## Metal to Semiconductor Transition in Niobium Sulfoselenide Alloy Films by Compositional Control

Tinsae Alem[1], Abir Hasan[2], Jordan A. Hachtel[3], Kory Burns[1], Nikhil Shukla[2], Stephen McDonnell[1]

[1]Department of Materials Science & Engineering, University of Virginia, Charlottesville, VA, USA
[2]Department of Electrical & Computer Engineering, University of Virginia, Charlottesville, VA, USA
[3]Center for Nanophase Materials Sciences, Oak Ridge National Laboratory, Oak Ridge, TN, USA

## Abstract

Transition metal dichalcogenides (TMDs) have garnered significant interest due to their tunable bandgaps, optical anisotropy, and mechanical flexibility. This study explores the relationship between stoichiometry, structural phase, and electronic transport in niobium sulfoselenide ($NbS_xSe_{2-x}$) thin films. In this study, low-dimensional TMD films were grown with molecular beam epitaxy (MBE) to investigate the effects of varying chalcogen (sulfur and selenium) content in $NbS_xSe_{2-x}$ alloys. We focus on tuning electrical resistivity along with the transition from metallic-to-semiconducting behavior through compositional control. These MBE grown films were characterized using in-vacuo x-ray photoelectron spectroscopy (XPS) to analyze the chemical composition. Electrical resistivity of the films and their temperature coefficient of resistance (TCR) were extracted to determine their metallic and semiconducting behavior. Our results show that increasing sulfur content induces a transition from metallic to semiconducting behavior in $NbSe_2$, accompanied by a rise in resistivity. This work highlights the potential for compositional tuning of TMDs for applications in tunable nanoelectronics.

## Introduction

Two-dimensional (2D) materials, have attracted significant attention in recent years due to their unique intrinsic properties, which set them apart from conventional three-dimensional (3D) materials.[1,2] These distinct characteristics often emerge when materials are scaled down to a single or few atomic layers, giving rise to electronic, optical, and structural behaviors not present in their bulk counterparts.[3,4] The rapid expansion of this field has been driven not only by seminal work with graphene[5–7] but also by the isolation of many other 2D materials; this includes transition metal dichalcogenides (TMD),[8] black phosphorus, and layered oxides. Collectively, these van der Waals (vdW) solids highlight the engineering of electronic and optical properties at the atomic scale.[3,4] The novel characteristics of 2D materials positions them as promising candidates for applications; including photodetectors, interconnects, transistors, and chemical sensors.

Among these classes of materials, TMDs have attracted an increasing scientific interest in a range of electronic platforms due to their versatile electrical, optical, physical, chemical, and mechanical properties.[2] TMDs are composed of a layer of transition metal sandwiched between two chalcogen layers, with the formula $MX_2$ where M is the transition metal and X is a chalcogen (e.g. M = Ti, Hf, Nb, Ta; X = S, Se, and Te). The planes between layers are bonded together by weak vdW forces, while the atoms are held tightly together in-planes by covalent bonds.[9] The weak interlayer interactions facilitate exfoliation into ultrathin layers, whereas the strong in-plane covalent bonding and absence of dangling bonds ensure structural stability in the monolayer form.

A variety of techniques have been used to synthesize TMDs, including molecular beam epitaxy (MBE), vapor-phase chemical reactions, chemical vapor deposition, wet-chemical synthesis, and liquid-phase exfoliation techniques.[10–13] MBE is valued among other techniques due to its ultra-high purity material growth that allows us to produce high-quality films.[14] This makes MBE ideal for creating thin films with controlled layer thickness, doping, and heterostructures; which are critical for advanced applications in optoelectronics, quantum devices, and fundamental materials research. [13] Such controlled growth is essential for studying phase stability, defect formation, and alloying behavior.

$NbSe_2$ is well known for its metallic conductivity, with a room-temperature resistivity of $(1.1 \pm 0.1) \times 10^{-4}$ Ω-cm,[15] and for its charge density wave (CDW) phase transition at low temperatures. In its bulk form, it exhibits a CDW transition around 33 K and becomes superconducting below 7.2 K.[16] $NbS_2$ is also reported as metallic, with a room-temperature resistivity of $(1.1 \pm 0.1) \times 10^{-4}$ Ω-cm,[15,17] similar to $NbSe_2$. However, $NbS_2$ can also exhibit semiconducting behavior depending on its phase, particularly in three-rhombohedral (3R) polytype. In addition, its electrical properties are sensitive to deviations from ideal stoichiometry, where the S/Nb ratio varies from the ideal 2:1.[18,19] Previous studies have shown that stoichiometry plays a critical role in controlling the electronic response of 3R-$NbS_2$: the 3R phase exhibits semiconducting behavior in the presence of Nb self-intercalation, while remaining metallic when self-intercalation is suppressed.[20,21]

The processing window for these phases varies widely across synthesis methods. For example, chemical vapor deposition (CVD) typically forms the 3R- $NbS_2$ between ~775-950˚C and 2H- $NbS_2$ is partially overlapping 840-930˚C range, while chemical vapor transport (CVT) studies have shown that 3R stabilization occurs under lower sulfur pressure and lower temperature (650˚C).[22,23] These observations indicate that $NbS_2$ is sensitive to growth conditions, stoichiometry, and chalcogen chemical potential. MBE, with its capability for precise flux control, therefore, provides a powerful platform for systematically exploring phase stability in $NbS_2$ and related alloys.

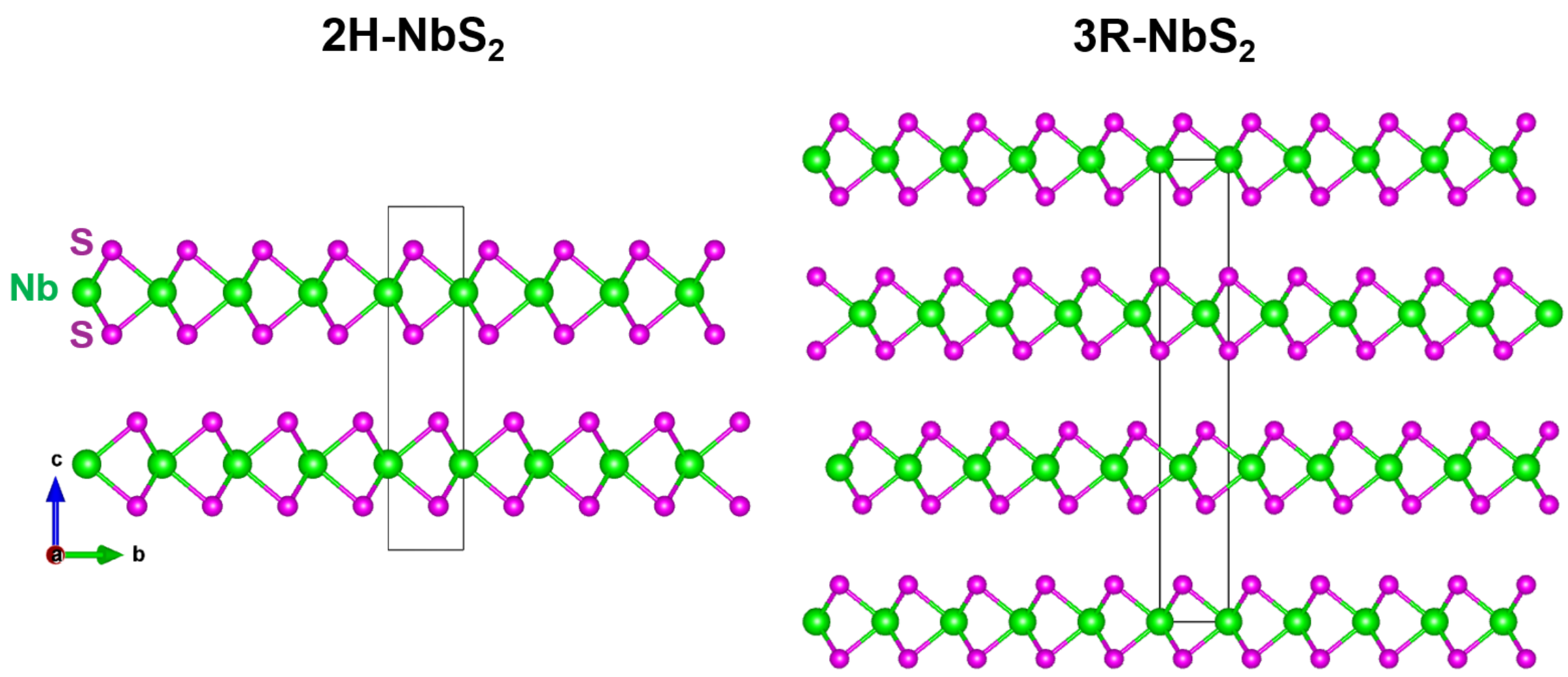

**Figure 1.** Crystal structures of 2H (hexagonal, two layers per unit cell; space group $P6_3/mmc$, no. 194) and 3R (rhombohedral, three layers per unit cell; space group *R3m*, no. 160) for $NbS_2$. Side views show Nb (green) sandwiched between two sulfur atoms (purple).

Although $NbS_2$ and $NbSe_2$ have been individually studied, comparatively little is known about the alloy system $NbS_xSe_{2-x}$.[24–26] NbSSe and related compositions offer the opportunity to continuously tune electronic structure through chalcogen substitution due to the lattice and electronic structure sensitivity to chalcogen size and bonding environment, whereas the binary system $NbS_2$ and $NbSe_2$ lack such compositional tunability. Understanding this behavior is critical because it provides insight into the interplay between stoichiometry, phase, and electronic properties in TMDs more broadly. In this work, we synthesize $NbS_xSe_{2-x}$ alloy thin films with $NbS_2$ and $NbSe_2$ as the two end-point compositions using MBE, while systematically tuning the S:Se flux ratio. By combining electrical transport measurements with structural analysis, we investigate how chalcogen composition influences the structural phase, stoichiometry, and ultimately the electronic properties of these alloys. Our results reveal a clear metal-to-semiconductor transition driven by sulfur incorporation. Notably, we show that this behavior is closely connected to Nb-rich stoichiometry and the stabilization of the 3R polytype. This study provides a controlled and unified picture of how compositional engineering can modulate the electronic ground states in Nb-based TMDs.

## Experimental

### MBE Growth

The $NbS_xSe_{2-x}$ films were grown using an MBE chamber described elsewhere.[27,28] The 1cm x 1cm p-doped silicon wafer substrates, 285 nm thick $SiO_2$/Si were cleaned with acetone, rinsed with isopropanol, and blown dry with nitrogen gas. The substrate was attached to a molybdenum sample holder by spot welding, two tantalum foils across opposite corners to secure the sample. Next, the sample was exposed to UV-$O_3$ for five minutes to remove residual contamination from the surface of the substrate before being introduced to the ultra-high vacuum (UHV) chamber (base pressure ~$1x10^{-9}$ mbar). Air exposure time between UV-$O_3$ and insertion was kept to a minimum and never exceeded 2 minutes. Once the substrate was in the UHV chamber, it was heated to ~250˚C and left overnight to outgas the surface before growth. Just prior to growth, the sample was heated to 800 ˚C to further outgas and clean the sample surface, while the Nb, S, and Se sources were also heated in preparation for deposition. Once the sample reached 800˚C and was held for 10 minutes, the temperature was then lowered to the growth temperature of 600˚C. A Fluke Endurance Series Pyrometer (optical pyrometer) was used to measure the sample temperatures above ~350˚C.

A beam flux monitor was utilized to measure beam equivalent pressure (BEP) to determine relative amount of metal and chalcogen impinging on the surface during growth. The experiment was carried out by varying the S and Se BEP while keeping the Nb BEP similar for every growth (Nb BEP ~$0.9x10^{-9}$ mbar). A Knudsen effusion cell was used for heating the pyrite ($FeS_2$) to evaporate S similar to a prior report.[29] $FeS_2$ at 430˚C-450˚C whereas the Se was evaporated using thermal cracker cell (TCC), with the TCC cracker held at 900˚C and the TCC reservoir temperature kept at 85˚C - 87˚C. An electron beam evaporator was used to heat the Nb source and is controlled by emission current and voltage. The sample shutter was closed when getting the metal and chalcogen ready for growth condition.

Once the sources were stabilized at the desired BEP, the S, Se, and Nb shutters were opened sequentially within ~2 seconds to initiate co-deposition onto the substrate. The growth progress was monitored in real time using reflection-high energy electron diffraction (RHEED). Each sample was grown over a period of ~5 hours with a growth rate of 2-3 layers/hour. After the growth was completed, the Nb shutter was closed, and the sample was held under the S and/or Se sources for two minutes at the growth temperature. After this, the manipulator sample heater was turned off, and the growth temperature gradually decreased to approximately 240˚C under S and Se fluxes. Cooling under a chalcogen flux helps prevent the film from decomposing into Nb metal. Finally, once the sample temperature reached ~240˚C, the S and Se shutters were closed, and the sample was allowed to cool down to room temperature naturally.

**X-ray Analysis**

The films were characterized by utilizing in-vacuo and ex-situ techniques. In-vacuo x-ray photoelectron spectroscopy (XPS) using an Al K alpha source (1486.7 eV) and an R3000 analyzer operated at 50 eV pass energy was used to analyze the chemical composition of the $NbS_xSe_{2-x}$ thin films. The XPS data were analyzed and fitted using AAnalyzer software. In addition, resistivity was calculated for each sample by carrying out an ex-situ measurement of the films sheet resistance with a Jandel 4-point probe, and considering its thickness, which was obtained by X-ray reflectivity (XRR) using a Rigaku SmartLab system. GSAS-II software[30] was used to fit the XRR data and estimate the film's thickness. X-ray diffraction (XRD) data were collected using an Empyrean diffractometer, with out of plane symmetry, using Cu K alpha radiation.

## Electron Microscopy

For additional structural characterization, $NbS_2$ samples were analyzed by electron microcopy. Here, the cross-sectional specimens for transmission electron microscopy (TEM) were prepared using a focused ion beam (FIB) in a dual-beam Helios UC G4 (Thermo Fisher Scientific). To minimize damage during FIB preparation, a protective $Al_2O_3$ capping layer was deposited prior to milling. Of note, sheet resistance was measured before both the capping and cross-sectional preparation steps. Scanning transmission electron microscopy (STEM) imaging was performed at an accelerating voltage of 200kV, with both bright-field (BF) and high-angle annular dark-field (HAADF) images acquired over identical regions. Core-loss electron energy loss spectroscopy (EELS) was performed using a Nion aberration-corrected UltraSTEM 100 operating at an accelerating voltage of 100 kV. The experiments were performed with a convergence semiangle of 30 mrad. A 2 mm EELS aperture was used to maximize the analytical signal during acquisition with an EELS collection semiangle of ~48 mrad.

## Electrical Measurement

Temperature coefficient of resistance (TCR) measurement was performed using Cascade Microtech Summit12000 and parameter were analyzed using Keithley 4200A-SCS. These films are prone to oxidation when removed from the vacuum system and exposed to ambient conditions for extended periods. A detailed study of this oxidation is the focus of ongoing work. To prevent this, samples intended for TCR measurements were coated with an $Al_2O_3$ protective layer using atomic layer deposition (ALD) within 10 minutes of air exposure.

## Results

To investigate the effect of chalcogen composition on electronic transport in TMDs, we synthesized a series of $NbS_xSe_{2-x}$ thin films using MBE. By systematically varying the sulfur and selenium fluxes, we achieved continuous tuning of the S:Se ratio while maintaining similar growth conditions. This compositional engineering allowed us to explore how chalcogen substitution influences structural and electronic properties.

## Composition Control and Growth

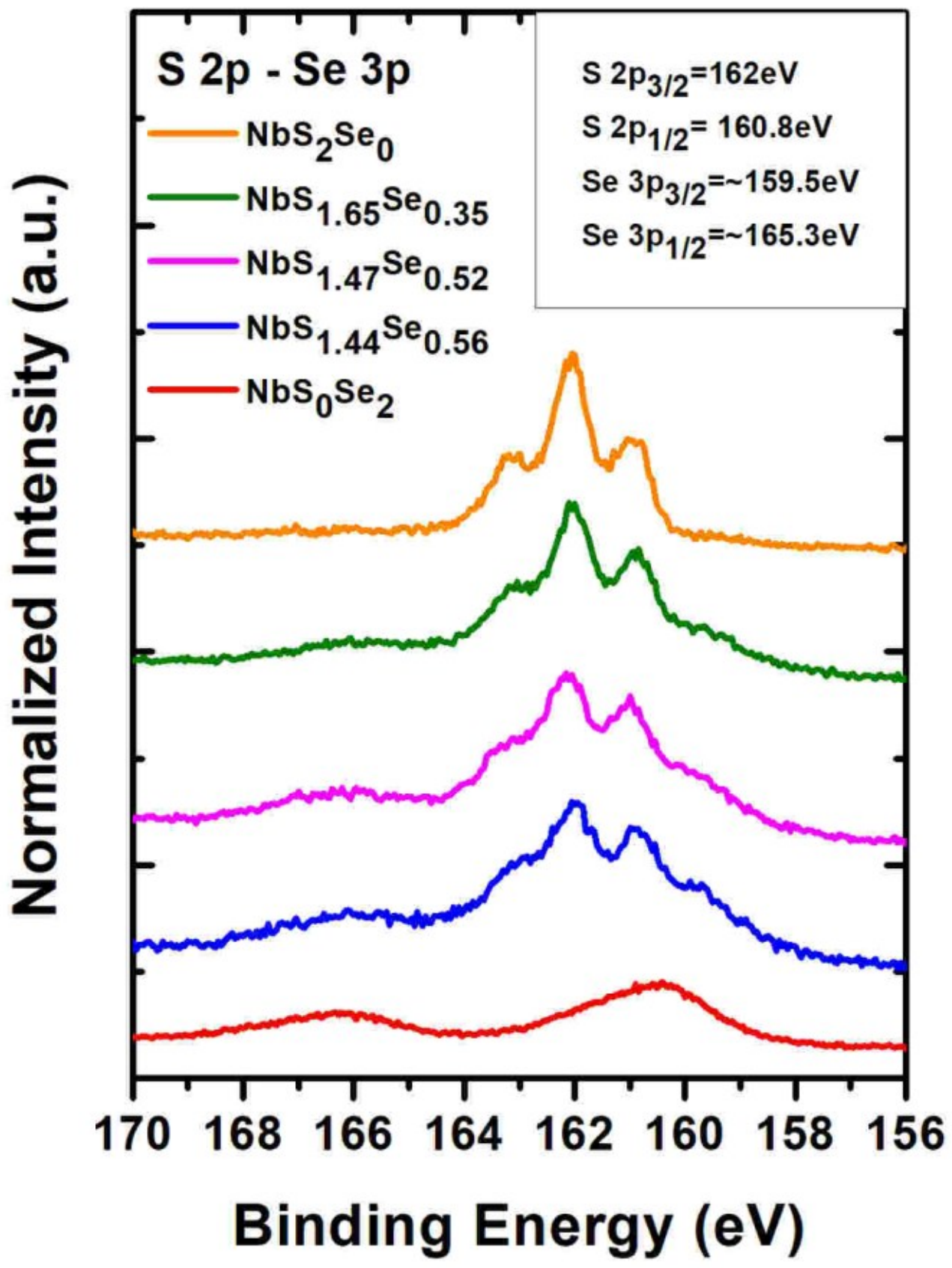

**Figure 2.** XPS spectra of $NbS_xSe_{2-x}$ for S 2p and Se 3p core level.

Figure 2 shows the XPS spectra of five $NbS_xSe_{2-x}$ compositions grown on a $SiO_2$/Si substrate, where the sulfur and selenium content were systematically varied. Figure 2 highlights the 156-170eV binding energy region, where both S 2p and Se 3p core levels are located. High-resolution scans in this region allow for the identification of sulfur and selenium contributions. The Se 3p doublet is expected to show a spin-orbit splitting of 5.7eV between the $3p_{3/2}$ (~160eV) and $3p_{1/2}$ (166eV) peaks, while the S 2p doublet shows a separation of 1.18eV, with the $2p_{3/2}$ peak appearing around 162eV.

For the $NbS_0Se_2$ sample shown in red, only the Se 3p doublet is present, consistent with the absence of sulfur in this sample. The subsequent spectra (blue, magenta, and green), representing intermediate $NbS_xSe_{2-x}$ compositions, show distinct peaks from both sulfur and selenium. Notably, the S $2p_{3/2}$ peak appears near 162 eV, while the Se $3p_{3/2}$ remains at ~160 eV. In the final spectrum (orange color, $NbS_2Se_0$), sulfur peaks are observed with only negligible Se peaks, again consistent with the absence of substantial selenium during growth. These observations confirm the successful control of chalcogen stoichiometry across the series.

**Electronic Properties**

The impact of chalcogen composition on electronic transport was determined by measuring the room temperature sheet resistance of $NbS_xSe_{2-x}$ films. Immediately after removing from vacuum system with less than 5 minutes of air exposure using a four-point probe. Subsequently, X-ray reflectivity (XRR) data was acquired using a Rigaku diffractometer, and the film thickness was estimated from this data and fitted using GSAS[30]. The electrical resistivity of the film was then calculated as the product of the sheet resistance and the film thickness. Figure 3 displays a plot of resistivity versus sulfur content. A clear trend appears as sulfur content increases that resistivity rises significantly. While Nordheim's rule for solid solutions was expected, we can see that the peak resistivity is not significantly higher than that for the $NbS_2$.

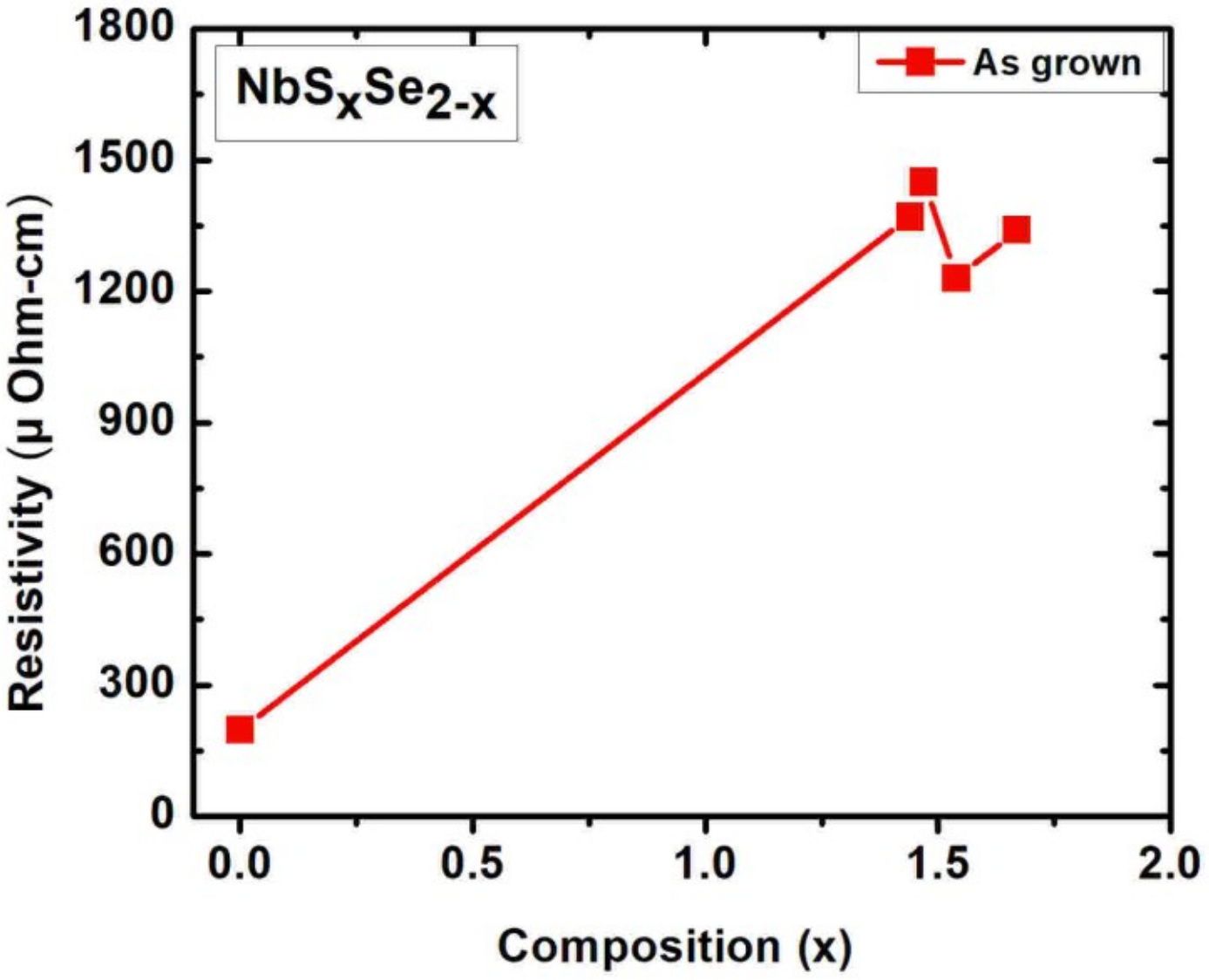

**Figure 3.** Electrical resistivity of $NbS_xSe_{2-x}$ as a function of sulfur content (x). The resistivity was calculated from measured sheet resistance and film thickness. $NbSe_2$ exhibits the lowest resistivity while $NbS_{1.47}Se_{0.53}$ shows significantly higher values.

$NbSe_2$ has a resistivity of 1.98E02 μΩ-cm. In contrast, $NbS_2$ exhibited resistivity values over an order of magnitude higher (1.34E03 μΩ-cm) than our $NbSe_2$ film, and higher than previously reported metallic $NbS_2$.[15] Intermediate compositions ($NbS_{1.44}Se_{0.56}$, $NbS_{1.47}Se_{0.53}$, and $NbS_{1.54}Se_{0.35}$) showed resistivities of 1.37E03, 1.45E03, and 1.23E03 μΩ-cm, respectively. To verify reproducibility, $NbSe_2$ films were grown five additional times and $NbS_2$ three additional times, and all measurements fell within a consistent range from 1.98E02 to 6.66E02 μΩ-cm for $NbSe_2$ and from 1.21E03 to 1.72E03 μΩ-cm for $NbS_2$. Representative films for each composition are shown in Figure 3, and the replicated data are provided in the Supplementary Information. The thickness of each grown film and the corresponding sheet resistance are also included in Supplementary Table 1-3.

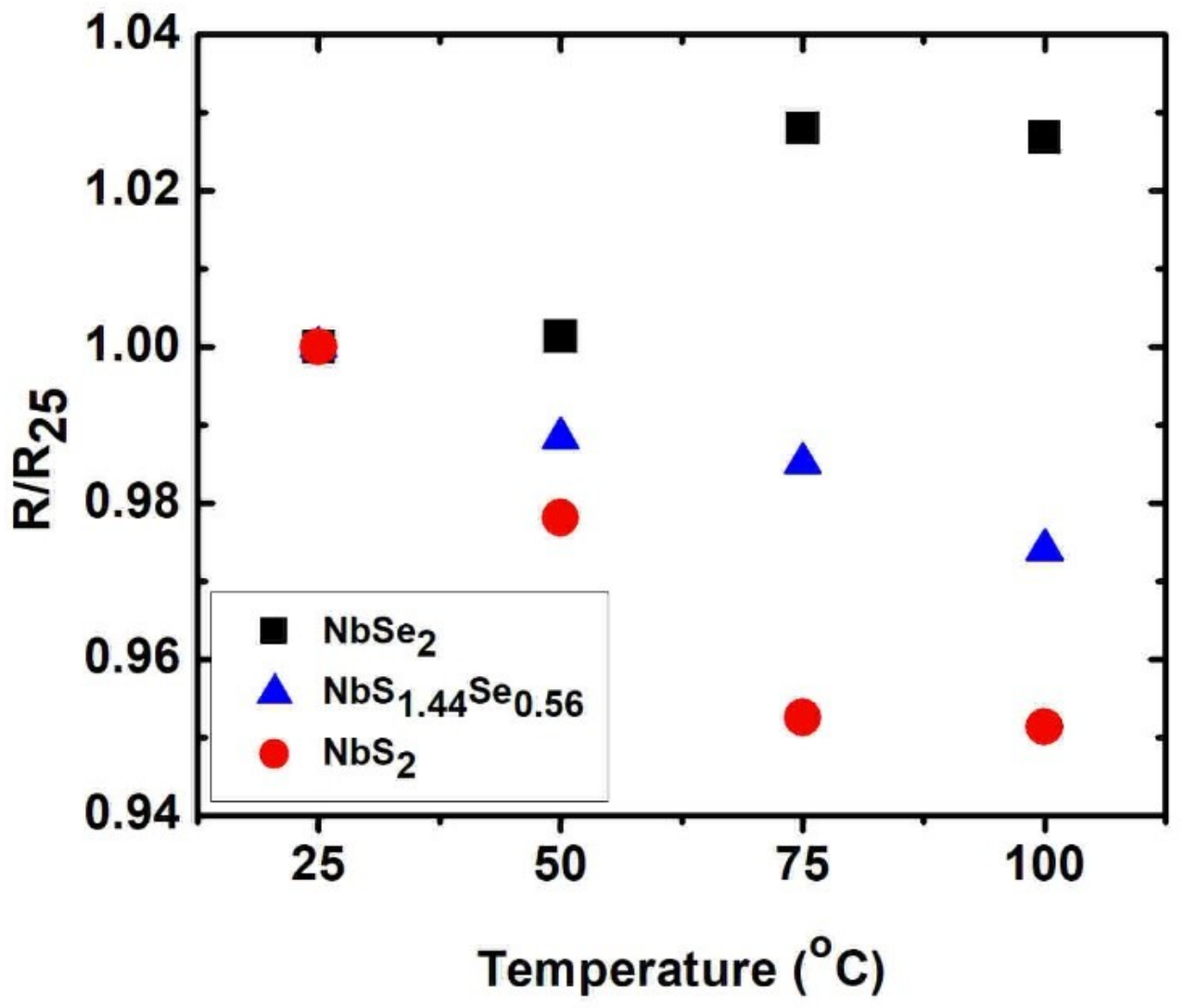

**Figure 4.** TCR measurements of $NbSe_2$, $NbS_{1.44}Se_{0.56}$, $NbS_2$. Resistivity is plotted as a function of temperature. $NbSe_2$ (metallic phase) shows a positive slope due to increased electron-phonon scattering; as a metal, it has free electrons that experience more frequent collision at higher temperatures. In contrast, $NbS_{1.44}Se_{0.56}$, $NbS_2$ (semiconducting phase) show a negative slope because the number of charge carriers increases with temperature, lowering resistivity.

Temperature dependent resistivity measurements (Figure 4) provide some insight into why the $NbS_2$ shows comparable resistivity to the intermediate compositions. As can be seen, while the $NbS_0Se_2$ shows a positive slope of resistivity as a function of temperature, consistent with metallic behavior, both the intermediate and $NbS_2Se_0$ samples show negative slopes of resistivity as a function of temperature, indicating semiconducting behavior.

**Structural and Chemical Analysis**

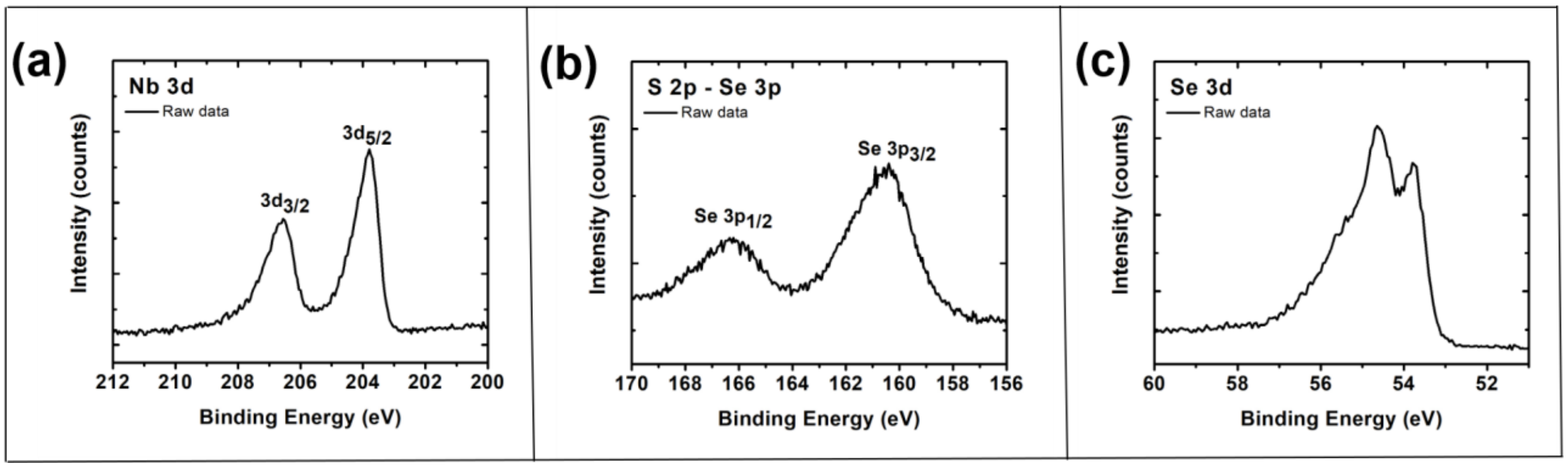

**Figure 5.** XPS of $NbSe_2$. (a) Nb 3d, (b) S2p-Se3p region but only Se 3p is present, and (c) Se 3d core level.

Figure 5 shows the Nb 3d and S 2p spectra for the $NbSe_2$ film. For $NbSe_2$, S 2p peak is absent, while the Se 3p doublet are present at 160eV ($3p_{3/2}$) and 166eV ($3p_{1/2}$) with 5.776 eV separation. The Se 3d region shows additional features that caused a broader peak shown in Figure 5c, consistent with self-intercalation, although the impact on electronic properties is not the same as for $NbS_2$.

Because $NbS_2$ can crystallize in either the 2H or 3R polytype, each with different stacking arrangements and electronic characteristics, we performed XPS and XRD analysis to determine whether the semiconducting behavior observed in our films (when sulfur is incorporated to the $NbSe_2$) is related to phase selection or stoichiometry.

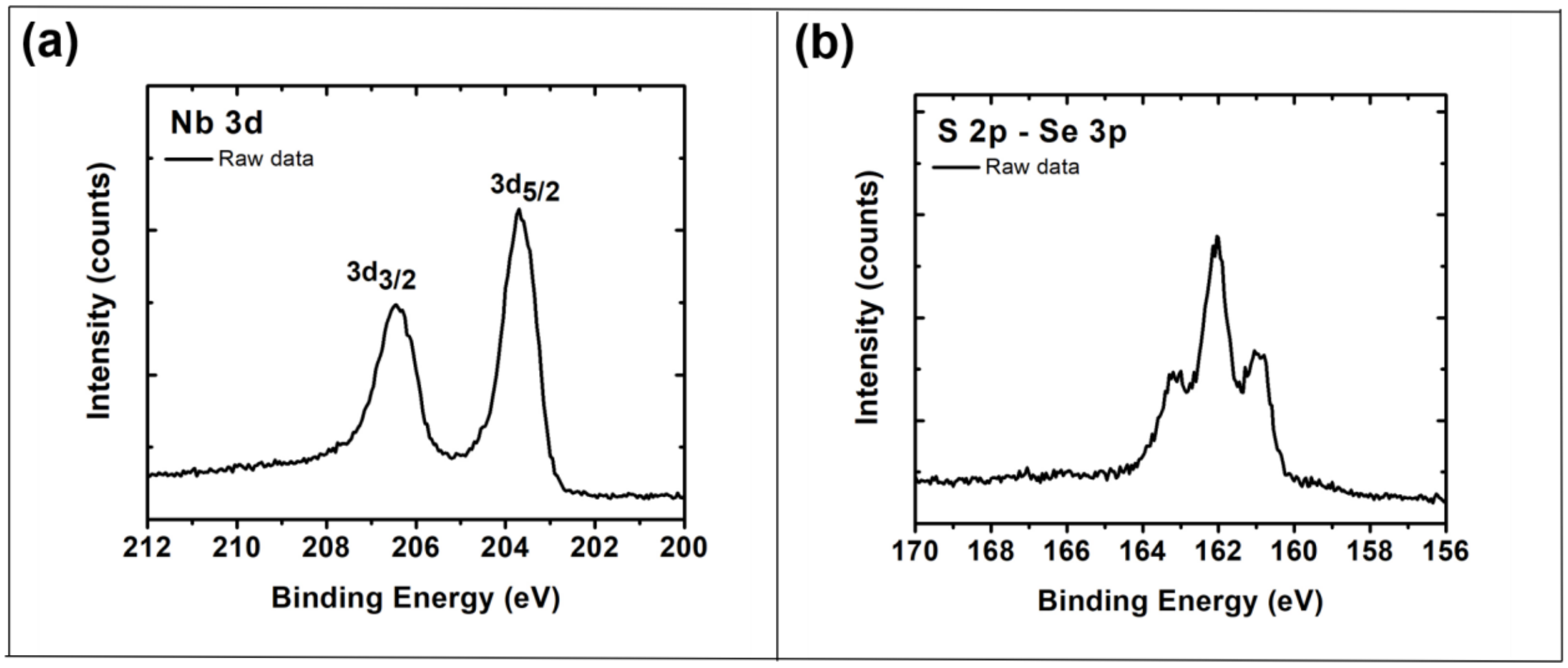

**Figure 6.** XPS spectra of $NbS_2$ grown for 5 hours: (a) Nb 3d and (b) S 2p.

Figure 6 shows the Nb 3d and S 2p spectra for the $NbS_2$ film. As expected for pure $NbS_2$ growth, no significant Se signal should be present; the weak features at 159.5 and 165.3 eV correspond to trace Se from the MBE chamber and are negligible. The S 2p region shows the expected doublet between 160-163 eV but also contains an additional higher binding energy doublet. Similar additional features and a chalcogen peak shifts have been reported in Nb-rich $NbSe_2$ and in V-rich $VTe_2$, where excess metal atoms self-intercalate into the vdW gap.[28,31] It is hypothesized that this intercalation modifies the electrostatic potential at the chalcogen sites due to the intercalation chalcogen-site being surrounded by more cations, resulting to shift the chalcogen core-level binding energies. Quantitative XPS of our film yields an S/Nb ratio of 1.67, indicating sulfur deficiency and Nb-rich, consistent with the presence of self-intercalated Nb.

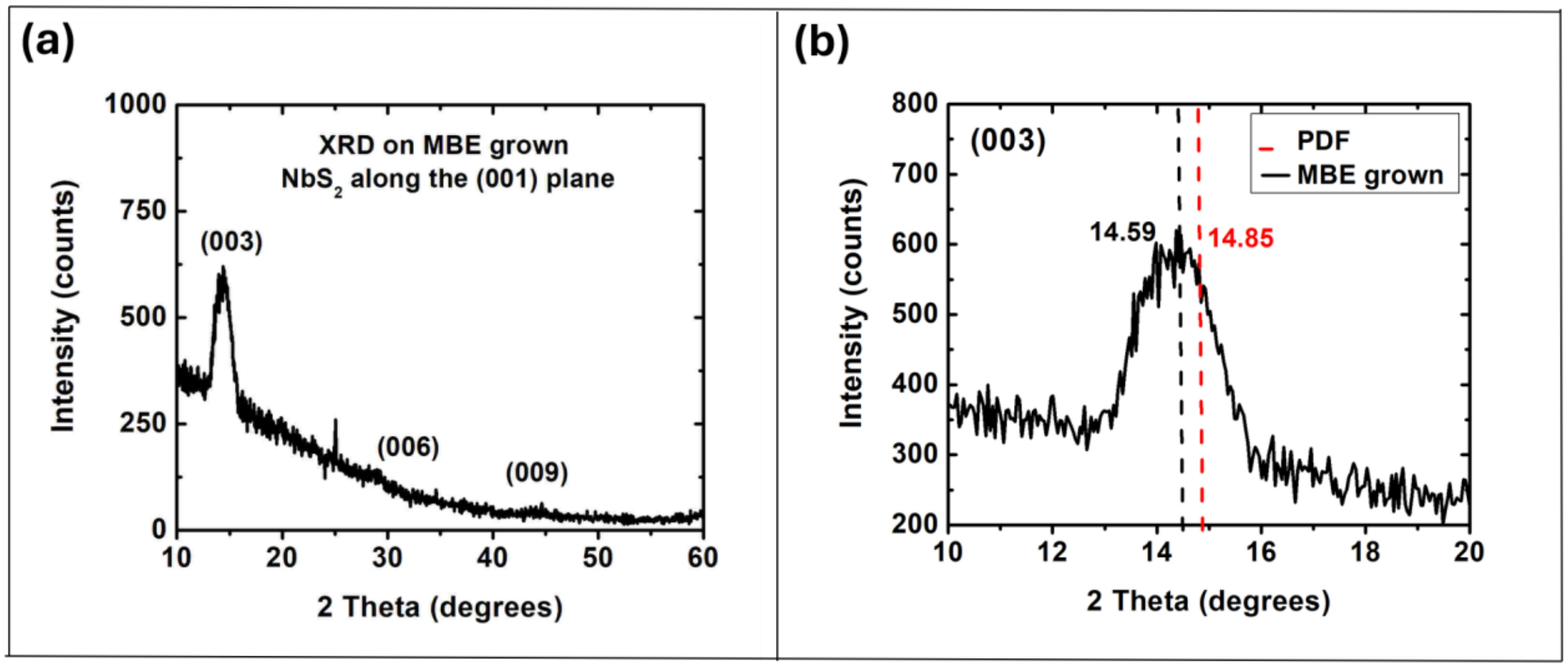

**Figure 7.** XRD analysis of as-grown 3R-$NbS_2$: (a) Full-range XRD scan from 10 to 60 2θ. (b) Magnified view from 10˚ to 20˚, highlighting the (003) reflection. The black dashed line indicates the 2θ positions from our MBE-grown $NbS_2$ film, while the red dashed line corresponds to the standard reference 2θ positions from the 3R-$NbS_2$ PDF (PDF 00-038-1367).

XRD measurements further support this interpretation, as shown in Figure 7. Although the 2H and 3R polytypes cannot be unambiguously distinguished using (00l) reflection alone, the c-axis lattice parameter provides important insight. The strongest peak in the 00l reflection for the $NbS_2$ and $NbSe_2$ films corresponds to 002 for the 2H polytype and 003 for the 3R polytype owing to their different layer stacking along c-axis. The reference 2 theta values for stoichiometric 2H or 3R are expected to be around 14.85˚, but in our data, the peak appears at 14.59˚ (with the first 00l peak, attributed to 003, shifted to lower angles, indicating an increased c-lattice parameter.) The standard 3R-$NbS_2$ reference (PDF 00-038-1367) lists a c-lattice parameter of 17.92Å (interlayer spacing of 5.97 Å).[22,23,32,33] In contrast, our $NbS_2$ film exhibits an expanded c-parameter of 18.20Å (interlayer spacing of 6.08 Å). Such expansion is a characteristic signature of self-intercalation, which increases the interlayer spacing by occupying the vdW gap. Reports indicate that this self-intercalation, which increases the interlayer spacing, is more thermodynamically favorable in the 3R phase than in the 2H phase due to its ABC stacking sequence. This is attributed to the 3R-phase being more readily available to accommodate excess metal atoms.[33]

## Discussion

These findings indicate that resistivity in $NbS_xSe_{2-x}$ can be tuned by changing the chalcogen composition, which in turn affects both the phase and the stoichiometry of the films. $NbSe_2$ (x=0, no sulfur) remains metallic, whereas sulfur incorporation increases the room-temperature resistivity and produces a negative TCR slope, indicative of semiconducting behavior. Sulfur substitution modifies the local bonding and electronic environment, significantly influencing transport properties. It is possible that with the correct choice of composition a flat TCR could be obtained within this temperature range.

The observed changes in transport can be explained by the formation of a Nb-rich, self-intercalated 3R-$NbS_2$ phase. Resistivity and XPS results strongly support the formation of a Nb-rich, self-intercalated 3R-$NbS_2$ phase. This is consistent with previous reports by Niazi et al. and Huang et al.,[20,22] which conclude that 3R-$NbS_2$ is non-stoichiometric and exhibits an order-of-magnitude lower conductivity due to excess interlayer Nb atoms donating electrons to the $dz^2$ conduction band and raising the Fermi level.[19] Self-intercalation arising from sulfur deficiency therefore plays a crucial role in $NbS_2$ films; the non-stoichiometric 3R-$NbS_2$ phase accommodates excess Nb, as confirmed by XPS and expanded c-axis lattice parameters in XRD. These donor-type interstitials reduce the electronic density of states near the Fermi level, diminishing carrier mobility.[21]

Further support for this comes from post-growth treatments in which the sulfur content of $NbS_2$ and intermediate Nb sulfoselenide films is increased. These treatments suppress Nb self-intercalation and lead to a corresponding improvement in electrical transport, as illustrated in Figure 8.

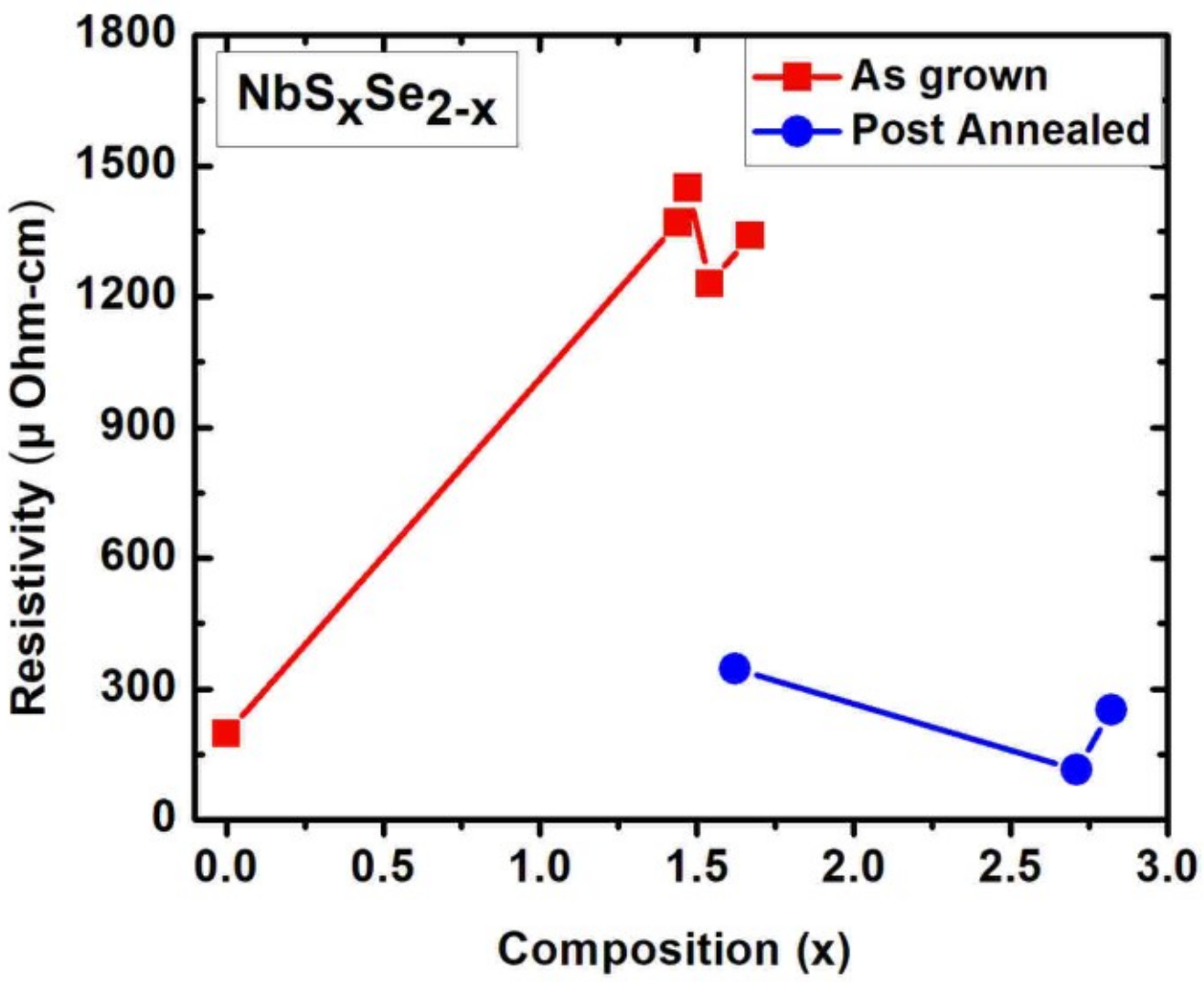

**Figure 8.** Resistivity of as grown and post annealed $NbS_xSe_{2-x}$ films plotted as a function of sulfur composition (x). Red symbols correspond to as grown films, while blue symbols represent post-annealed samples.

Figure 8 summarizes the resistivity as a function of composition for both as grown and post annealed $NbS_xSe_{2-x}$ films. As grown $NbS_2$ film with an S/Nb ratio of 1.67, exhibiting a resistivity of 1.34E03 μΩ-cm. After annealing an identically grown film under sulfur flux at 800˚C for 18.5hr, the S/Nb ratio increases to 2.71, and the resistivity decreased to 1.14E02 μΩ-cm. This order of magnitude indicates that mitigating sulfur deficiency suppresses Nb self-intercalation and improves electrical transport. Also, XPS measurements on a separate sample provides additional evidence for this trend. The S/Nb ratio increases from 1.66 in the as-grown film to 2.82 after annealing at 800˚C for 3hrs under sulfur flux. Resistivity measured after annealing is 2.53E02 μΩ-cm, again substantially lower than that of sulfur deficient as-grown $NbS_2$.

Despite the increased S/Nb ratios exceeding the ideal stoichiometric ratio of 2 in the post-annealed samples, their XRD patterns exhibit similar diffraction features as as-grown $NbS_2$, suggesting that sulfur annealing does not produce an obvious change in the overall crystal structure detectable by XRD (Figure 9a). $NbS_3$ and $NbS_2$ are expected to exhibit distinct diffraction patterns. In particular, $NbS_3$ is expected to exhibit three main diffraction peaks at 2 theta values of approximately 10˚, 19.8˚, and 40˚, corresponding to the (001), (002), and (004) planes, respectively.[34–36] The absence of these characteristic peaks in the XRD patterns of our MBE-grown niobium sulfide further suggests that $NbS_3$ did not form (Figure 9a). The possibility that the elevated S/Nb ratios arise from sulfur accumulation at the surface is also considered; however, no elemental sulfur (binding energy at 164 eV)[37] is detected in the post-annealed niobium sulfide, as shown in Figure 9b. Nevertheless, given the uncertainties in quantitative XPS analysis, particularly those associated with sensitivity factors, an estimated ~20% relative compositional uncertainty is considered when interpreting the measured S/Nb ratios.[38–41]The measured ratios are reported to facilitate comparison between samples, rather than to establish absolute stoichiometries.

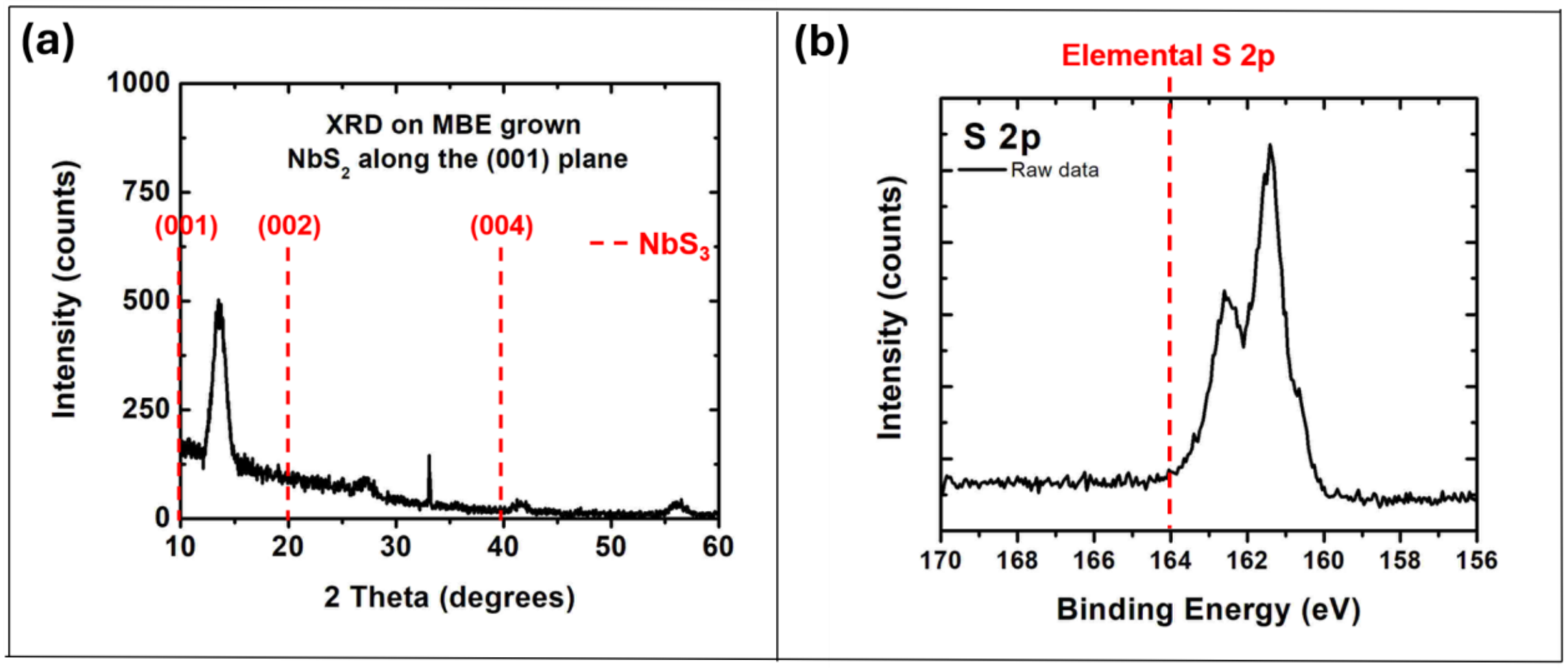

**Figure 9.** Structural and chemical analysis of post annealed $NbS_2$. (a) XRD pattern of MBE-grown $NbS_2$, with the red dashed line indicating the reference position of the $NbS_3$ peak. (b) S 2p XPS spectrum of the same sample, showing peaks attributed to Nb-S bonding rather than elemental sulfur. The red dashed line indicates the expected position of the elemental S 2p signal for reference.

The intermediate sulfoselenide, NbSSe, was also evaluated after post-growth treatment, which the as-grown sample was annealed under simultaneous S and Se flux. A similar trend is observed, with the treatment significantly improving conductivity and reducing the resistivity to 3.46E02 μΩ-cm, approximately an order of magnitude lower, as shown in Figure 8. In addition to this work, an alternative approach was also explored.

Here, an intermediate sulfoselinde sample was synthesized via a conversion method, in which $NbS_2$ is first grown and annealed under sulfur flux to increase the sulfur content, followed by annealing under selenium flux to form NbSSe. In this approach, the initial anneal under sulfur flux ensures that the $NbS_2$ reaches a sulfur content at which semiconducting behavior associated with sulfur deficiency is no longer observed prior to alloy formation. This approach also leads to a significant improvement in conductivity, yielding a resistivity of 3.32E02 μΩ-cm, corresponding to an approximately one order of magnitude reduction compared to 1.27E03 μΩ-cm of as grown NbSSe. These results suggest that chalcogen mediated suppression of self-intercalation plays a key role in driving the semiconductor-to-metal transition in the NbSSe alloy system.

Further analysis was carried out for both as-grown (Figure 10 a and b) and post annealed (Figure 10 c and d) $NbS_2$ using aberration-corrected STEM for detailed local structural analysis. Two separate samples were prepared for this purpose: one in the as-grown state and the other subjected to a post-growth sulfur annealing treatment. The sheet resistance of each sample was first measured to confirm consistency with previously characterized as grown and post-annealed samples. Cross-sectional STEM imaging was then performed to examine the layer structure of each sample, as shown in Figure 10. In both samples, the amorphous $Al_2O_3$ capping layer is visible on top of the $NbS_2$ films, followed by the $NbS_2$ layers, while part of the amorphous $SiO_2$ substrate is visible beneath the $NbS_2$. The observed layered structure and step features provide direct measure of the $NbS_2$ film thickness. For the as-grown sample (Figure 10 a, b), the observed step features indicate seven stacked layers, consistent with van der Waals bonded $NbS_2$, corresponding to a film thickness of 4.9nm (assuming a monolayer thickness of 0.7nm.[42,43]) The measured sheet resistance of 2.92E9 μΩ/sq yields a calculated resistivity of 1.43E3 μΩ-cm, consistent with values obtained for other as-grown samples. For the post-annealed $NbS_2$ (Figure 10 c, d) the sheet resistance was measured as 1.71E8 μΩ/sq before cross-sectional STEM imaging. STEM analysis reveals a thickness of ~9.8nm, corresponding to 14 layers. This yields a calculated resistivity of 1.68E2 μΩ-cm, consistent with other post-annealed samples, which show a dramatic, order-of-magnitude decrease in resistivity.

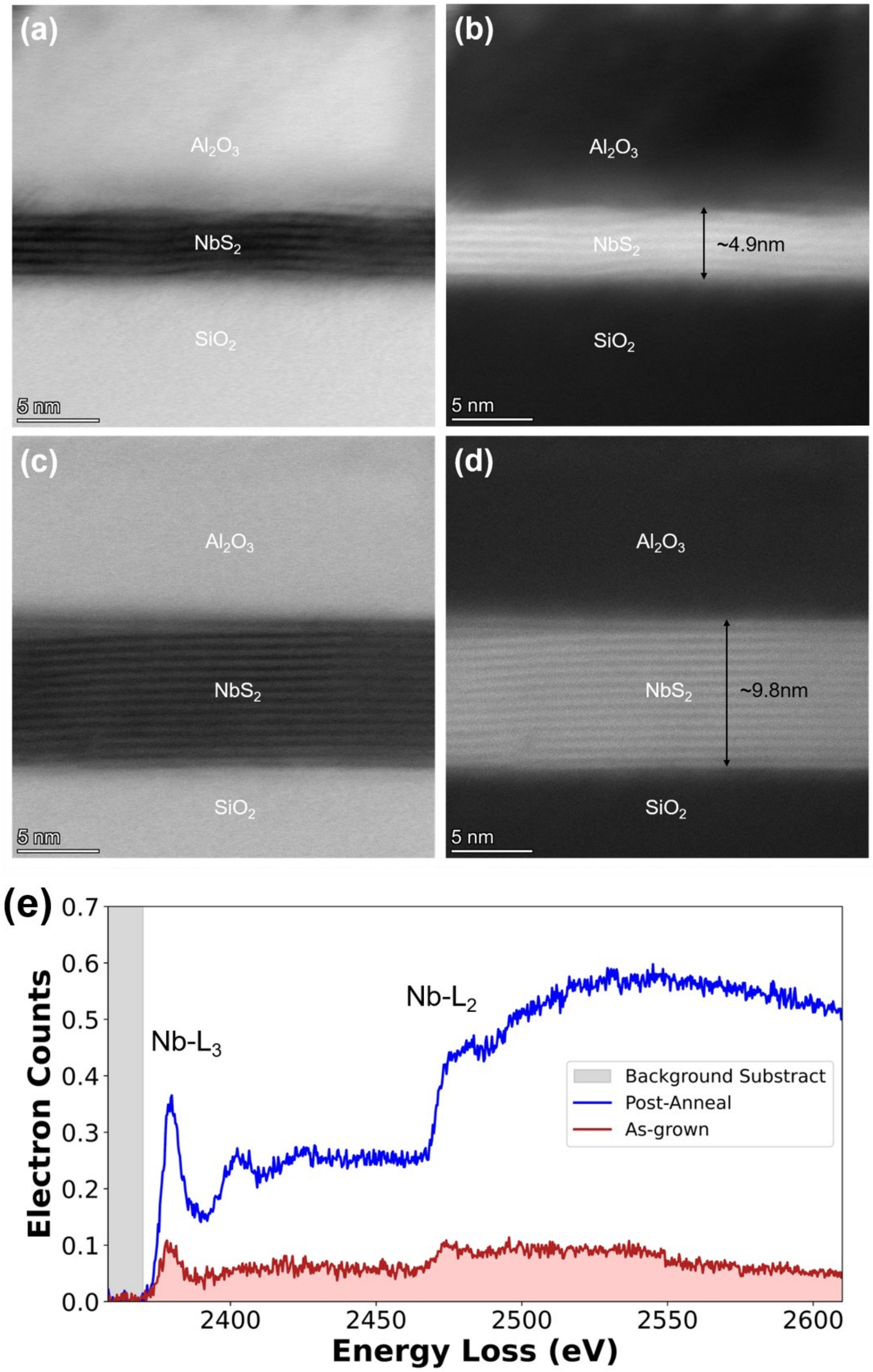

**Figure 10.** Cross-sectional STEM images of two separately prepared $NbS_2$ samples, consisting of an as-grown sample (a, b) and a post annealed sample (c, d), acquired at 3.8Mx magnification. Panel (a) shows bright field (BF) and (b) high-angle annular dark-field (HAADF) of as grown $NbS_2$. Panel (c) displays the BF image, while (d) shows the HAADF image of the post annealed sample, showing the collection of low-angle and high-angle scattered electrons, respectively. The different layers are labeled accordingly. At this scale, the amorphous $Al_2O_3$ capping layer is visible on top, followed by the $NbS_2$

layers. Beneath the $NbS_2$, part of the amorphous $SiO_2$ substrate is also visible. (e) EELS spectra of the Nb $L_{2,3}$ edges of the as-grown and post annealed $NbS_2$ samples. The red spectrum corresponds to the as-grown sample, while the blue spectrum represents the post-annealed sample. The $L_3$ and $L_2$ edges, arising from Nb $2p_{3/2}$ and $2p_{1/2}$ core-level transitions to unoccupied 4d states, respectively, are indicated.

Figure 10c and d shows that the post-annealed $NbS_2$ layers are uniform and well-defined, with even interlayer spacing indicating continuous out-of-plane stacking. In contrast, the as-grown layers have a “wavy” like periodicity (shown in Figure 10 a and b), suggesting minor structural perturbations. The intercalation of atoms into layered TMDs is known to modify the interlayer interactions and can lead to structural heterogeneities. This includes altered stacking, local strain, or modified interlayer spacing of the host lattice. While our aberration-corrected STEM imaging does not directly resolve atomic scale distortions, the discontinuous (or “wavy”) appearance of the as-grown (self-intercalated 3R-$NbS_2$) may result from intercalation-induced perturbations of the layered structure. These distortions appear to be reduced following post-growth sulfurization as shown in Figure 10 c and d).

Figure 10e shows the STEM-EELS spectra in the Nb L-edge region, where the red and blue spectra correspond to the as-grown and post-annealed $NbS_2$, respectively. The Nb $L_3$ edge shifts from 2375.32 eV in the as-grown sample to 2376.21 eV in the post-annealed sample. Qualitatively, this shift toward high energy loss suggests a modification of the local electronic environment of Nb, potentially associated with changes in coordination and oxidation state following sulfurization. In the as-grown $NbS_2$, self-intercalated Nb atoms may contribute to a modified local electronic environment, whereas the increased sulfur content following post-annealing likely alters the local Nb coordination and electronic structure.

Overall, self-intercalation affects $NbSe_2$ less strongly than $NbS_2$, as chalcogen deficiency in $NbS_2$ alters phase stability and electronic properties. While the rhombohedral (ABC) stacking may slightly reduce interlayer orbital overlap, the primary factor driving semiconducting behavior appears to be the combination of interstitials and sulfur-induced localization of Nb d-electrons.[21] These effects explain why $NbS_2$ and intermediate NbSSe films exhibit semiconducting behavior. Collectively, these observations demonstrate a clear link between chalcogen composition, structural phase, stoichiometry, and electronic behavior. Finally, compositional engineering controls phase stability and Nb intercalation, which in turn dictates the electronic ground state in these layered TMD alloys.

## Conclusions

Our results indicate that the introduction of sulfur induces a transition from metallic-to-semiconducting behavior. The stabilized phase of our $NbS_2$ films was identified as a non-stoichiometric form of 3R-$NbS_2$. Furthermore, we observed a clear trend of increasing resistivity with increasing sulfur content in the as-grown thin films. These findings demonstrate that both stoichiometry and crystal phase play a critical role in determining the electronic transport properties of TMD alloys. Our results underscore the potential of compositional tuning as a strategy for engineering TMDs with tailored electronic characteristics such as flat TCR behavior. In addition, this work paves the way for the realization of their application in next generation nanoelectronics devices such as transistors, interconnects, and semiconductors. Future investigations, including Hall effect and ARPES measurements might be essential for elucidating the underlying mechanisms responsible for the observed behavior.

## Acknowledgments

The authors acknowledge funding from Intel Corporation. This work was also supported by the Department of Materials Science and Engineering Olsen Fellowship. XRD, FIB, and STEM measurements were conducted at the Nanoscale Materials Characterization Facility (NMCF) at the University of Virginia. We appreciate the assistance of Dr. Helge Heinrich in training the author to use the Helios UC G4 Dual Beam FIB and the Themis TEM. Core-loss EELS conducted as part of a user project at the Center for Nanophase Materials Sciences (CNMS), which is a US Department of Energy, Office of Science User Facility at Oak Ridge National Laboratory.

Supplementary Information
to
Metal to Semiconductor Transition in Niobium Sulfoselenide Alloy Films by Compositional Control

Tinsae Alem[1], Abir Hasan[2], Jordan A. Hachtel[3], Kory Burns[1], Nikhil Shukla[2], Stephen McDonnell[1]
[1]Department of Materials Science & Engineering, University of Virginia, Charlottesville, VA, USA
[2]Department of Electrical & Computer Engineering, University of Virginia, Charlottesville, VA, USA
[3]Center for Nanophase Materials Sciences, Oak Ridge National Laboratory, Oak Ridge, TN, USA

**Growth Parameter and Resistivity**

The following Tables S1, S2, and S3 summarize the growth time, thickness, sheet resistance, and resistivity of the samples characterized in this study. Replicated samples of $NbSe_2$, NbSSe, and $NbS_2$ exhibit some sample-to-sample variation; however, the overall resistivity trends remain consistent.

Table S1. Growth parameters of $NbSe_2$, including growth time, thickness, and resistivity. Samples are labeled numerically (01-06) for ease of tracking. The thickness of each sample was estimated from XRR data and fitted using GSAS software, except for sample 02, for which the film thickness was determined using TEM (marked as *).

| Sample | Growth # | Sheet Resistance (μΩ/sq) | Growth time (hr) | Thickness (nm) | Resistivity (μΩ-cm) |
|---|---|---|---|---|---|
| $NbSe_2$ | 01 | 6.84E+08 | 5 | 6.5 | 4.45E+02 |
| | 02* | 1.14E+09 | 5 | 4 | 4.56E+02 |
| | 03 | 3.60E+08 | 5 | 5.5 | 1.98E+02 |
| | 04 | 9.25E+08 | 5 | 7.2 | 6.66E+02 |
| | 05 | 5.65E+08 | 5 | 7.9 | 4.46E+02 |
| | 06 | 6.67E+08 | 5 | 7.7 | 5.14E+02 |

Table S2. Growth parameters of NbSSe (intermediate samples), including growth time, thickness, and resistivity. Samples are labeled numerically (07-10) for ease of tracking. The thickness of each sample was estimated from XRR data and fitted using GSAS software.

| Sample | Growth # | Sheet Resistance (μΩ/sq) | Growth time (hr) | Thickness (nm) | Resistivity (μΩ-cm) |
|---|---|---|---|---|---|
| NbSSe | 07 | 3.96E+09 | 1.75 | 3.1 | 1.23E+03 |
| | 08 | 1.96E+09 | 5 | 7.4 | 1.45E+03 |
| | 09 | 1.76E+09 | 5 | 7.8 | 1.37E+03 |
| | 10 | 1.44E+09 | 4 | 8.8 | 1.27E+03 |

Table S3. Growth parameters of $NbS_2$, including growth time, thickness, and resistivity. Samples are labeled numerically (11-14) for ease of tracking. The thickness of each sample was estimated from XRR data and fitted using GSAS software, except for growth 14, where TEM was used to estimate the thickness, marked as * (see Figure 9 in the main text).

| Sample | Growth # | Sheet Resistance (μΩ/sq) | Growth time (hr) | Thickness (nm) | Resistivity (μΩ-cm) |
|---|---|---|---|---|---|
| $NbS_2$ | 11 | 1.07E+10 | 5 | 1.6 | 1.72E+03 |
| | 12 | 2.57E+09 | 3 | 5.2 | 1.34E+03 |
| | 13 | 2.38E+09 | 3.5 | 5.1 | 1.21E+03 |
| | 14* | 2.92E+09 | 4 | 4.9 | 1.43E+03 |

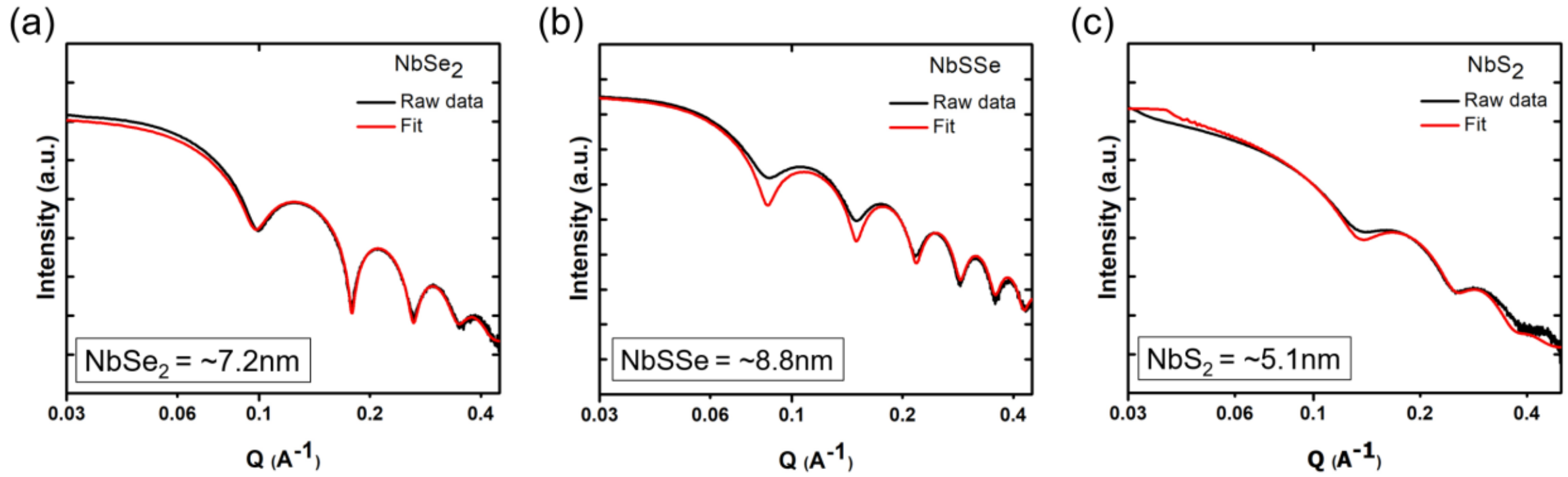

Figure S1. Representative X-ray reflectivity (XRR) spectra of $NbSe_2$ (a), NbSSe (b), and $NbS_2$ (c) grown on $SiO_2$/Si substrates, it's showing the intensity as a function of the scattering vector, Q ($Å^{-1}$). Panel (a) shows $NbSe_2$ from growth 04 with a thickness of 7.2nm; panel (b) shows NbSSe from growth 10 with a thickness of 8.8nm; and panel (c) corresponds to $NbS_2$ from growth 13 with a thickness of 5.1 nm; this is the same sample shown in Figure 7 of the main text, where XRD spectra are presented.

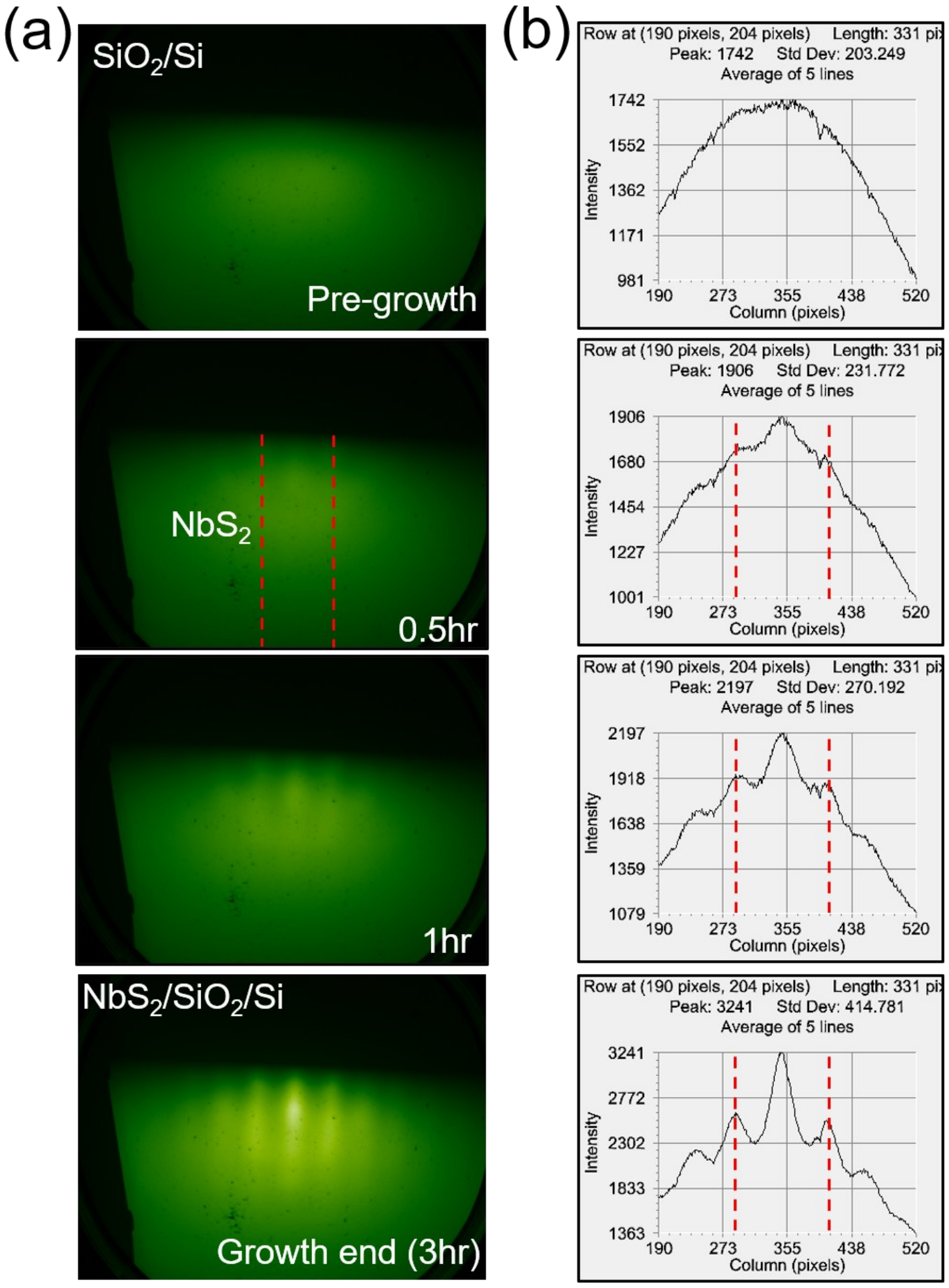

Figure S2. RHEED patterns of $NbS_2$ during growth. (a) RHEED images acquired at the pre-growth stage and after 0.5, 1, and 3 hr of growth (top to bottom). (b) Corresponding line profiles extracted from the RHEED patterns (top to bottom). The red dashed lines indicate the expected reciprocal lattice features of $NbS_2$. No distinct features are observed in the pre-growth pattern due to the amorphous $SiO_2$ substrate. With increasing growth time (starting from 0.5hr), new diffraction features emerge, indicating the formation of the first $NbS_2$ monolayer. These features become more pronounced in both the RHEED patterns and the corresponding line profiles, where additional side peaks and a higher-intensity central peak develop. The progressive sharpening and increased intensity of these peaks reflects improved film ordering and crystallinity.

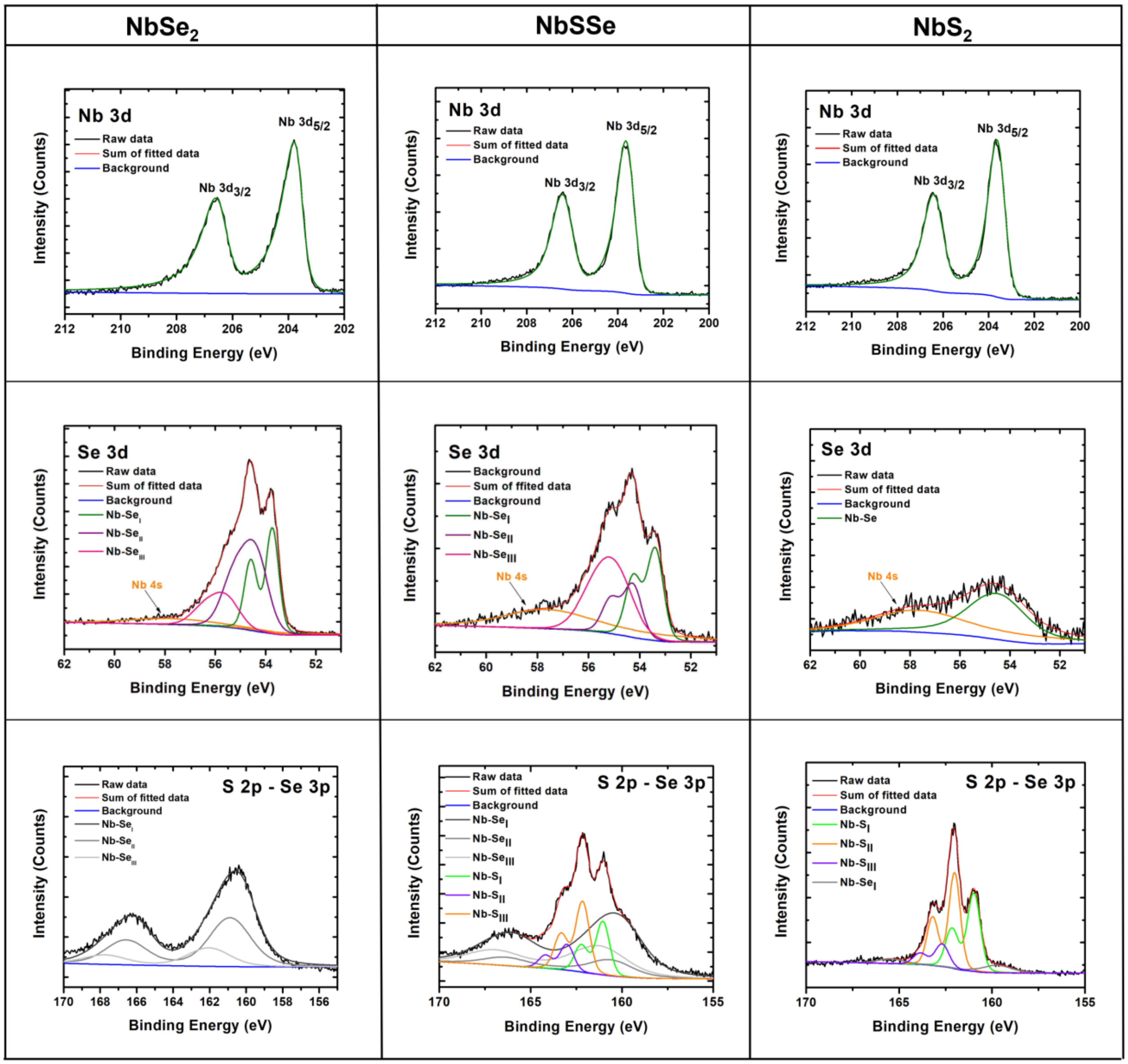

Figure S3. X-ray photoelectron spectroscopy (XPS) spectra of as-grown $NbSe_2$, NbSSe, and $NbS_2$. The corresponding Nb 3d, Se 3d, and S 2p-Se3p core-level spectra are presented to characterize the evolution of the chemical states across this composition. In-vacuo XPS measurements were performed using monochromatic Al K alpha source (1486.7 eV) with a pass energy of 50eV to analyze the chemical composition and chemical states of NbSSe samples. The sensitivity factors used for quantitative analysis were 4.86 and 3.35 for Nb $3d_{5/2}$ and Nb $3d_{3/2}$, respectively; 1.11 and 0.567 for S $2p_{3/2}$ and S $2p_{1/2}$, respectively; and 1.36 and 0.934 for Se $3d_{5/2}$ and Se $3d_{3/2}$, respectively.[1] The relative sensitivity factors for the Se 3p components were 5.744 and 2.989, respectively, and were determined experimentally in our laboratory using elemental Se as a reference. The spectra were analyzed and fitted using Aanalyzer software.

**Post Growth Treatment effects**

Table S4 summarizes the annealed samples, including detailed information on thickness, sheet resistance, and resistivity. The post-growth treatment was performed without exposing the samples to air; therefore, the thickness was estimated using XRR after annealing. Growth 16 (post-annealed $NbS_2$) from this table is shown in Figures S5 and S6, which present the RHEED pattern and the XRR fitting plot, respectively.

Table S4. Summary of the effects of post-growth annealing on $NbS_2$ and NbSSe. The cumulative annealing time is also included. The thickness of all samples listed here was extracted from XRR fitting, except for growth 17, where STEM imaging was used to estimate the thickness (see Figure 10 in the main text).

| Sample | Growth # | Cumulative post growth anneal time (hr) | Anneal Under chalcogen flux | Sample Thickness (nm) | Sheet of Resistance (μΩ/sq) | Resistivity (μΩ-cm) |
|---|---|---|---|---|---|---|
| $NbS_2/SiO_2/Si$ | 15 | 3 | S | 6 | 4.22E08 | 2.53E02 |
| | 16 | 18.5 | S | 5.1 | 2.25E08 | 1.14E02 |
| | 17* | 18.3 | S | 9.8 | 1.71E08 | 1.68E02 |
| $NbSSe/SiO_2/Si$ | 18 | 5 | S and Se | 8.7 | 3.98E08 | 3.46E02 |
| | 19 | 6 | S and Se | 6.2 | 5.36E08 | 3.32E02 |

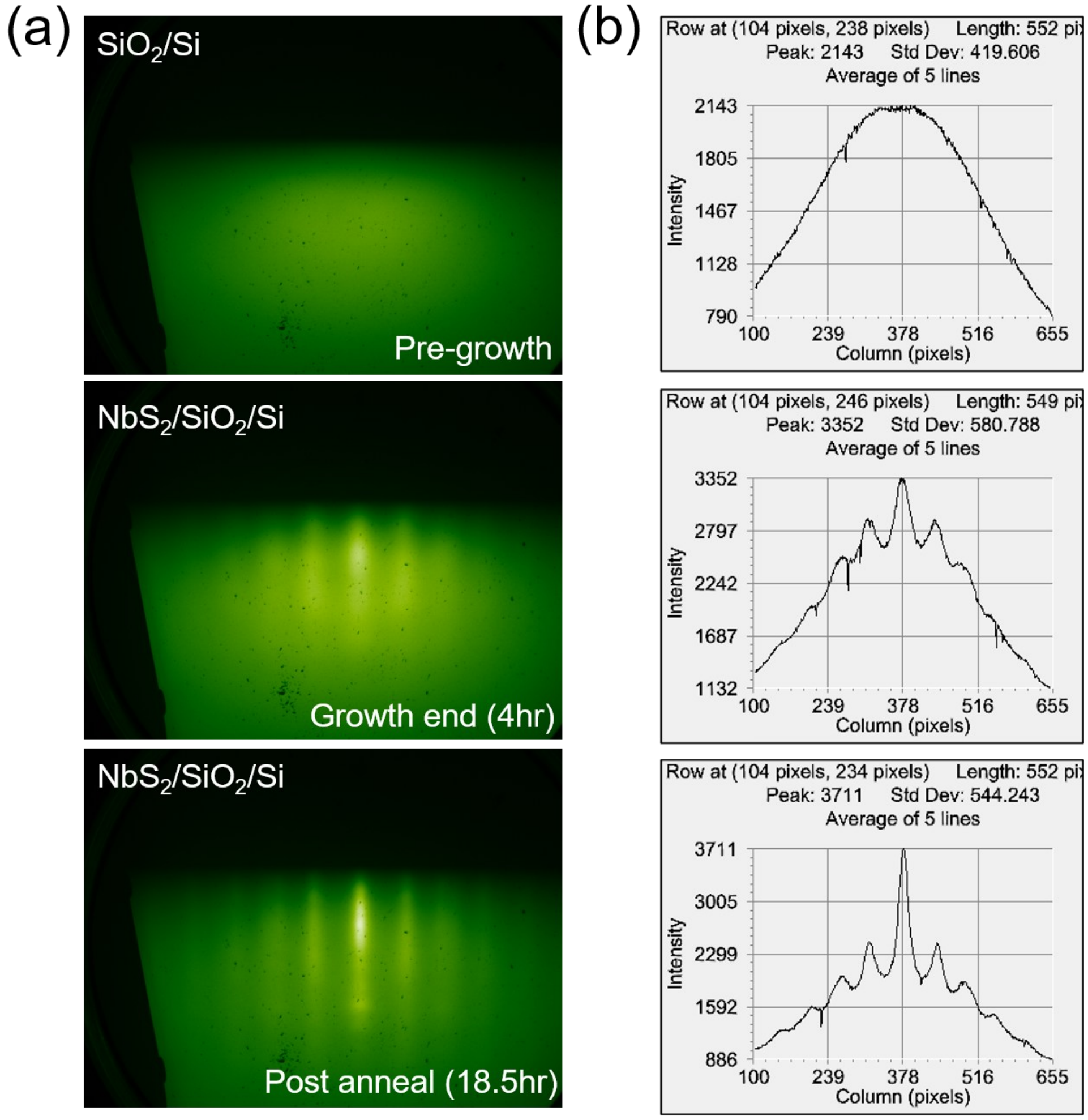

Figure S5. Post-growth effects are further illustrated through RHEED patterns of $NbS_2$ grown on a $SiO_2$/Si substrate, Growth 16. (a) RHEED images acquired at the pre-growth stage and after 4hr of growth and 18.5hr post anneal (top to bottom). (b) Corresponding line profiles extracted from the RHEED patterns (top to bottom). The top panel shows the pre-growth stage, where only the amorphous $SiO_2$ substrate is visible. The middle panel corresponds to the film after 4 h of $NbS_2$ growth. The bottom panel shows the sample after 18.5hr of annealing under sulfur flux at 800C. The RHEED pattern becomes progressively sharper after annealing, showing improved crystallinity (although the RHEED was acquired at different temperature).

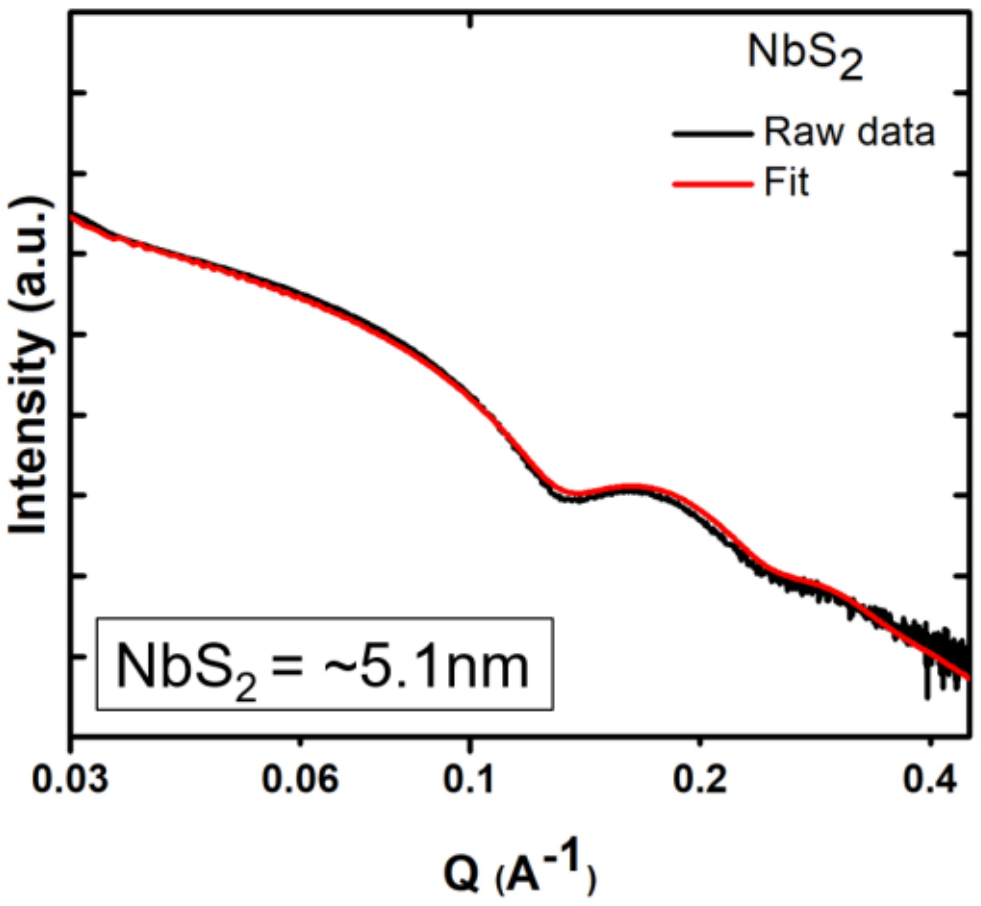

Figure S6. X-ray reflectivity (XRR) of post-annealed $NbS_2$ grown on a $SiO_2$/Si substrate. This dataset corresponds to the $NbS_2$ sample presented in Figure 8 of the main text and is related to the sample shown in Figure S5 of this Supplementary Information. The sample was annealed for 18.5 hours. Analysis of the XRR data yields an estimated $NbS_2$ film thickness of 5.1 nm.